\documentclass[final]{l4dc2025}

\def\isarxiv{0}

\title[Toward Near-Globally Optimal 
Nonlinear Model Predictive Control via Diffusion Models]{Toward Near-Globally Optimal 
Nonlinear Model Predictive Control via Diffusion Models
\if\isarxiv1
-- Extended Version
\fi
}
\usepackage{times}
\usepackage{amsmath} 
\usepackage{amssymb}  
\usepackage{bm}
\usepackage{graphicx}
\usepackage{algorithm}
\usepackage[noend]{algorithmic}
\usepackage[capitalise]{cleveref}
\usepackage{nicefrac}
\usepackage{pgfplots}
\usepackage{tikzviolinplots}
\usepackage{hyperref}
\usepackage{xurl}

\usetikzlibrary{arrows,shapes,backgrounds,patterns,fadings,matrix,arrows,calc,
	intersections,decorations.markings,
	positioning,external,arrows.meta}
\usetikzlibrary{pgfplots.statistics}
\usepgfplotslibrary{fillbetween}

\DeclareMathOperator*{\argmin}{arg\,min}

\newcommand{\changed}[1]{\textcolor{black}{#1}}

\author{%
 \Name{Tzu-Yuan Huang}$^{1*}$ 
 \Email{tzu-yuan.huang@tum.de}\\
 \Name{Armin Lederer} $^{2*}$
 \Email{armin.lederer@inf.ethz.ch}\\
 \Name{Nicolas Hoischen}$^{1}$ 
 \Email{nicolas.hoischen@tum.de}\\
 \Name{Jan Br\"udigam}$^{1}$ 
 \Email{jan.bruedigam@tum.de}\\
 \Name{Xuehua Xiao}$^{1}$ 
 \Email{xuehua.xiao@tum.de}\\
 \Name{Stefan Sosnowski} $^{1}$ \Email{sosnowski@tum.de}\\
 \Name{Sandra Hirche}$^{1}$ 
 \Email{hirche@tum.de}\\
$^{1}$\addr Chair of Information-oriented Control, Technical University of Munich, Germany\\
$^{2}$\addr  Learning and Adaptive Systems Group, Department of Computer Science, ETH Zurich, Switzerland\\
$^{*}$\addr Equal contribution
}

\pgfplotsset{
	legend cell align = left, ticklabel style = {font=\scriptsize},
	every axis label/.append style={font=\scriptsize},
	legend style = {font=\tiny},title style={yshift=-7pt, font = \scriptsize}, compat=1.3 }

\pgfplotsset{
compat=1.11,
legend image code/.code={
\draw[mark repeat=2,mark phase=2]
plot coordinates {
(0cm,0cm)
(0.19cm,0cm)        
(0.38cm,0cm)         
};%
}
}

\begin{document}

\setlength{\textfloatsep}{7pt}
\setlength{\floatsep}{4pt}

\setlength{\abovedisplayskip}{4pt}
\setlength{\belowdisplayskip}{4pt}

\maketitle

\begin{abstract}%
Achieving global optimality in nonlinear model predictive control (NMPC) is challenging due to the non-convex nature of the underlying optimization problem. Since commonly employed local optimization techniques depend on carefully chosen initial guesses, this non-convexity often leads to suboptimal performance resulting from local optima. To overcome this limitation, we propose a novel diffusion model-based approach for near-globally optimal NMPC consisting of an offline and an online phase. 
The offline phase employs a local optimizer to sample from the distribution of optimal NMPC control sequences along generated system trajectories through random initial guesses. Subsequently, the generated diverse dataset is used to train a diffusion model to reflect the multi-modal distribution of optima.  
In the online phase, the trained model is leveraged to efficiently perform a variant of random shooting optimization to obtain near-globally optimal control sequences without relying on any initial guesses or online NMPC solving.
The effectiveness of our approach is illustrated in a numerical simulation indicating high performance benefits compared to direct neural network approximations of NMPC and significantly lower computation times than online solving NMPC using global optimizers. 
\end{abstract}


\vspace{-0.2cm}
\section{Introduction}
\vspace{-0.15cm}

Nonlinear Model predictive control (NMPC) \citep{rawlings2017model} is a modern control technique that iterates between solving an optimal control problem and applying the first element of the solution's control sequence to the nonlinear system. Due to the ease of considering constraints and the flexibility for optimizing performance metrics, NMPC has immense success in a large variety of applications ranging from robot manipulators \citep{heins2023keep} and unmanned aerial vehicles \citep{lindqvist2020nonlinear} to chemical engineering \citep{drgovna2017optimal}.

The optimal control problems inherent in NMPC are generally non-convex, but commonly solved using gradient-based optimization techniques in practice \citep{rao1998application, qin1997overview}. However, such methods only ensure convergence to a local optimum close to the initial guess. The optimized cost function can reflect metrics such as energy consumption \citep{OLDEWURTEL201215}, agricultural output \citep{ding2018model}, and material consumption \citep{ghezzi2023implicit}. Therefore, poor local optima are suboptimal solutions that can have an immediate detrimental real-world effect. This has led to the development of globally optimal NMPC approaches, often relying on randomness to ensure sufficient exploration, e.g., particle swarm optimization \citep{7175048} and the cross-entropy method \citep{wang2019exploring, pinneri2021sample}. However, these approaches are computationally demanding, restricting their application to settings with low sampling times.\looseness=-1

Since NMPC is already considered computationally expensive without the integration of global optimization, different approaches have been proposed to reduce its online complexity. A straightforward approach is to reduce the number of iterations required by the numerical solver using warm starts \citep{yildirim2002warm}, which can be obtained from a function approximator trained from offline generated solutions \citep{10685132} or policies obtained via reinforcement learning \citep{hart2019lane,reiter2024ac4mpc}. While this approach still requires online optimization, explicit NMPC approaches aim at directly determining a function that solves the optimal control problem of NMPC \citep{bemporad2002explicit}. In recent years, it has become increasingly popular to achieve this by using flexible function approximators such as neural networks to learn the NMPC control law in a supervised fashion from numerically solved optimal control problems \citep{8371312,karg2020efficient,9006856,BONZANINI2021107174,tagliabue2024efficient}. Even though these approaches significantly reduce the complexity of NMPC, they do not consider the challenges of the multi-extremal optimization landscapes.

This issue has recently received increasing attention in the related field of trajectory planning and off-policy reinforcement learning, where generative models have shown promising results \citep{urain2024deep}. In particular, diffusion models are highly effective at learning multi-modal representations from sample data \citep{huang2023diffusion}. For trajectory planning, this can be leveraged to employ complex trajectory optimization techniques to generate expert data, whose distribution can be learned with a diffusion model to enable fast online planning \citep{janner2022planning, carvalho2023motion}. In offline reinforcement learning, diffusion models conditioned on normalized rewards can learn to predict the best state sequence from roll-outs of arbitrary policies jointly with a tracking controller \citep{ajay2023is}. Moreover, the full action distribution can be directly learned from demonstrations of the policy \citep{chi2023diffusion}, which can be used to warm start model predictive control \citep{li2024efficient} and for efficient sampling-based online optimization \citep{zhou2024diffusion}. However, these approaches focus on different challenges compared to the problem of learning a globally optimal NMPC controller: High diversity of samples is the major goal of diffusion models for planning, while in offline reinforcement learning the training data of diffusion models cannot be influenced.\looseness=-1

In this paper, we address the problem of finding a tractable approximation of globally optimal NMPC via diffusion models. For this purpose, we empirically approximate a multi-modal distribution over the optimal control sequences of NMPC in the form of a data set, which we obtain using local optimization methods that are optimized for low computation times. This distribution can be straightforwardly learned from the dataset using diffusion models, such that new samples can be drawn from it with comparatively little computational complexity. Finally, we exploit the model knowledge to employ a variant of the random shooting method to find control sequences close to the global optimum using samples from the diffusion model. The numerical results suggest that our approach based on learning a distribution over solutions strongly outperforms methods directly learning the control law while exhibiting lower complexity than an online global optimization strategy.\looseness=-1

\vspace{-0.2cm}
\section{Problem Formulation} \label{problem_statement}
\vspace{-0.15cm}

We consider a nonlinear discrete-time control system defined by differentiable system dynamics

\noindent$\bm{f}:\mathbb{R}^{n_x} \times \mathbb{R}^{n_u} \rightarrow \mathbb{R}^{n_x}$, i.e.,
    \begin{equation} \label{eq:system}
        \bm{x}_{t+1} = \bm{f}(\bm{x}_t, \bm{u}_t),
    \end{equation}
with time $t \in \mathbb{N}$, state vector $\bm{x}_t\in\mathbb{X}\subset\mathbb{R}^{n_x}$ in the compact set $\mathbb{X}$, and control input \changed{$\bm{u}_t\in \mathbb{U} \subset \mathbb{R}^{n_u}$}. 
We assume that the inputs $\bm{u}_t$ are constrained to some known compact set $\mathcal{U}\subset\mathbb{U}$.
This work aims to find an optimal input sequence $\bm{u}^*_{[0:H]}(\bm{x}_t)=\begin{bmatrix}
    \bm{u}^*_0(\bm{x}_t), \ldots, \bm{u}^*_{H-1}(\bm{x}_t)
\end{bmatrix}$ over a time horizon $H\in\mathbb{N}$, solving the following Optimal Control Problem (OCP) at every time step $t$ 
\begin{subequations} \label{eq:MPC_problem}
        \begin{align} 
            \underset{\bm{u}_{[0:H]}}{\min} J(\bm{x}_t, \bm{u}_{[0:H]}) &:= \sum_{i=0}^{H-1} \ell(\hat{\bm{x}}_i, \bm{u}_i) + V_f (\hat{\bm{x}}_H) \\
            \mathrm{s.t.} \quad &\hat{\bm{x}}_{i+1} = \bm{f}(\hat{\bm{x}}_i, \bm{u}_i),  \label{eq:MPC_dynamics} \; \hat{\bm{x}}_0 = \bm{x}_t, \\
            &\; \bm{u}_{[0:H]}=\begin{bmatrix}
                \bm{u}_0,\ldots,\bm{u}_{H-1}
            \end{bmatrix},\; \bm{u}_i \in \mathcal{U}, \; i=0, \ldots, H-1. \label{eq:MPC_constraint}
        \end{align}
    \end{subequations}
where the cost function $J:\mathbb{X}\times\mathbb{U}^H\rightarrow \mathbb{R}$
consists of the differentiable stage cost $\ell:\mathbb{X}\times\mathbb{U}\rightarrow \mathbb{R}$
and the differentiable terminal cost $V_f:\mathbb{X}\rightarrow\mathbb{R}$.
In the NMPC framework, given an initial state $\bm{x}_0$, the OCP \eqref{eq:MPC_problem} can be solved recursively over a moving horizon $[t, t+H]$ at each discrete sampling time instant $t$. This process generates an optimal input sequence $\bm{u}^*_{[0:H]}(\bm{x}_t)$ at every time step $t$, and only the first element $\bm{u}^*_0(\bm{x}_t)$ in this sequence is applied as the control input of the current time step.\looseness=-1 

While \eqref{eq:MPC_problem} generally constitutes a non-convex optimization problem for nonlinear dynamics \eqref{eq:system}, it is commonly solved numerically using gradient-based optimization techniques. Considering these numerical solvers as black boxes, they can be interpreted as operators $\mathbb{S}: \mathcal{C}^1(\mathbb{U}^H)\times \mathbb{U}^H\!\rightarrow\! \mathcal{U}^H$ that map the cost function $J(\bm{x}_t,\!\cdot)$ to a locally optimal solution $\hat{\bm{u}}_{[0:H]}(\bm{x}_t)$ which depends on the initial guess $\tilde{\bm{u}}_{[0:H]}$. Even though one of the locally optimal solutions $\hat{\bm{u}}_{[0:H]}(\bm{x}_t)$ corresponds to the globally optimal solution $\bm{u}_{[0:H]}^*(\bm{x}_t)$, we generally do not know which initial guess $\tilde{\bm{u}}_{[0:H]}$ needs to be chosen to obtain it. Therefore, we cannot expect to find the globally optimal control sequence $\bm{u}_{[0:H]}^*(\bm{x}_t)$ without further assumptions on the initial guess $\tilde{\bm{u}}_{[0:H]}$, such that we consider a slightly weaker notion of optimality in the following, which we refer to as probabilistic near-global optimality. We define a stochastic policy $\bm{\pi}:\mathbb{R}^{n_u}\times\mathbb{R}^{n_x}\rightarrow \mathbb{R}_{\geq 0}$ to satisfy this form of optimality, if\looseness=-1
\begin{align}\label{eq:control_goal}
    P(||\bm{u}- \bm{u}^*_{[0:H]}(\bm{x}_t)||_{2}\leq \varepsilon)\geq 1-\delta \text{ for } \bm{u}\sim\bm{\pi}(\cdot|\bm{x}_t)
\end{align}
for constants $\varepsilon,\delta\!\in\!\mathbb{R}_{\geq 0}$. The goal of this paper is the development of a method for finding such probabilistic near-globally optimal control laws $\bm{\pi}$ \changed{satisfying \eqref{eq:control_goal} with arbitrarily small $\varepsilon,\delta$} 
without relying on local online optimization.
\vspace{-0.2cm}

\begin{remark}
    Note that we do not consider state constraints in the OCP \eqref{eq:MPC_problem} because our focus lies in efficiently finding approximations of the globally optimal control sequence $\bm{u}^*_{[0:H]}(\bm{x}_t)$. Since the stochastic policy \eqref{eq:control_goal} does not inherit exact constraint satisfaction from the OCP \eqref{eq:MPC_problem} in general, extending it to state constraints is left for future work.
\end{remark}
\vspace{-0.4cm} 
\vspace{-0.1cm}
\section{Approximating Nonlinear Model Predictive 
Control via Diffusion Models} \label{main_method}
\vspace{-0.1cm}

We propose an approach for approximating globally optimal NMPC via diffusion models based on a procedure with an offline training phase and an online optimization phase.
In the offline training phase, we generate samples from a distribution over the optimal solutions of the NMPC problem \eqref{eq:MPC_problem} by exploiting computationally efficient local optimization methods. The details of our employed sampling scheme are explained in \cref{sec:data_generation}. 
These samples are used to train a diffusion model in \cref{sec:train}, which can approximate the distribution of optimal control sequences arbitrarily accurately given sufficient model flexibility and data set size. Therefore, generating a sufficiently large number of samples using the diffusion model will yield at least one control sequence close to the globally optimal one, which is exploited for efficient online optimization in \cref{sec:inference}.\looseness=-1


\vspace{-0.1cm}
\subsection{Data Generation for Diffusion-Approximated NMPC} \label{sec:data_generation}
\vspace{-0.1cm}

While the multi-extremal nature of NMPC's optimization landscape is well-known, it is not trivial to 
describe the corresponding solutions. Aligning this representation with our goal of training a diffusion model, we propose to use a probability distribution $\rho(\cdot|\bm{x}_t):\mathbb{U}^H\rightarrow \mathbb{R}$ over the control sequences, which peaks at the solutions of \eqref{eq:MPC_problem}. This distribution can be straightforwardly defined using the solution operator $\mathbb{S}$ and an uniformative, flat distribution $\phi:\mathbb{U}^H\rightarrow \mathbb{R}$ with support on $\mathcal{U}^H$, i.e., \looseness=-1
\begin{align}
    \mathbb{S}(J(\bm{x}_t,\cdot),\tilde{\bm{u}}_{[0:H]})\sim \rho(\cdot|\bm{x}_t)&& \text{for } \tilde{\bm{u}}_{[0:H]}\sim\phi.
\end{align}
Even though determining an explicit description of $\rho(\cdot|\bm{x}_t)$ is generally intractable, sampling from it merely requires sampling from $\phi$ and subsequently solving the OCP \eqref{eq:MPC_problem} numerically. This can be efficiently realized using local optimization methods designed for fast optimization in NMPC. Therefore, the distribution $\rho(\cdot|\bm{x}_t)$ can be effectively represented through samples in principle. 

As $\rho(\cdot|\bm{x}_t)$ is a conditional distribution dependent on the current state $\bm{x}_t$, it does not suffice to generate these samples for a single state. However, densely covering $\mathbb{X}$, e.g., by using a uniform distribution over $\mathbb{X}$, is also intractable for many problems due to the curse of dimensionality. Therefore, we design a distribution that puts an emphasis on states close to the trajectories generated under the locally optimal policies starting from a region of interest. This distribution relies on roll-outs of control inputs sampled from $\rho(\cdot|\bm{x}_t)$ starting from initial states $\bm{x}_0\sim \chi$ randomly drawn from an initial distribution $\chi:\mathbb{X}\rightarrow \mathbb{R}$, which is only supported on a comparatively small subset $\mathbb{X}_0\subset\mathbb{X}$. Since the employed policy in this procedure does generally not correspond exactly to the policy that will be used for the control of the actual system, this approach will experience the covariance shift phenomenon leading to compounding errors between trajectories in the data set and trajectories generated by the policy applied to the system \citep{rajaraman2020toward}. To mitigate this issue, we augment the sampling scheme by randomly perturbing states with Gaussian noise $\bm{\epsilon}\sim\mathcal{N}(\bm{0},\bm{\sigma}(t))$ with time-dependent variance $\bm{\sigma}^2:\mathbb{N}\rightarrow\mathbb{R}_{\geq 0}$. Thereby, we generate a data set that will asymptotically be dense in $\mathbb{X}$ and thus, inherits robustness against compounding errors from the uniform distribution, while exhibiting significant advantages for learning policies in the areas of interest defined via trajectories. The complete algorithm for sampling states and optimal sequences to generate a representative data set $\mathcal{D}$ of $\rho(\cdot|\bm{x}_t)$ is summarized in \cref{alg:data_generation}.

It is important to note that we only sample one control sequence $\hat{\bm{u}}_{[0:H]}(\bm{x}_t)$ per state $\bm{x}_t$ from the distribution of solutions $\rho(\cdot|\bm{x}_t)$, which merely requires the execution of a local optimization method to evaluate $\mathbb{S}$. This is in contrast to multistart techniques for global optimization that require multiple samples from $\rho(\cdot|\bm{x}_t)$ \citep{danilova2022recent}. However, it is straightforward to show that 
this approach asymptotically leads to a data set with that contains a dense subset of global optima.\looseness=-1
\if\isarxiv0%
\footnote{\label{fn:extension}Proofs for all results can be found in the extended version (\url{https://mediatum.ub.tum.de/doc/1779727}).\looseness=-1}%
\fi%

\vspace{-0.05cm}
\begin{theorem}\label{th:data}
    Consider data sets $\mathcal{D}$ generated using Algorithm~\ref{alg:data_generation} with positive $\bm{\sigma}$. If $P(\bm{u}_{[0:H]}^*(\bm{x}))>0$ for all $\bm{x}\in\mathbb{X}$ under the distribution induced by $\phi$ and $\mathbb{S}$, then, for every $\bm{x}\in\mathbb{X}$, $\epsilon\in\mathbb{R}$, $\bar{\delta}\in\mathbb{R}$, there exist $N_s,N_p,N_T\in\mathbb{N}$ such that $(\bm{x}',\bm{u}^*_{[0:H]}(\bm{x}'))\in\mathcal{D}$ and $||\bm{x}-\bm{x}'||\leq \epsilon$ with probability $1-\bar{\delta}$.
\end{theorem}
\if\isarxiv1
\begin{proof}
    Since $\bm{\sigma}$ is positive, the probability density of $\bm{x}_t^d$ is supported everywhere on $\mathbb{X}$. Hence, the probability of the set $\{\bm{x}': ||\bm{x}-\bm{x}'||\leq \epsilon\}$ under the distribution of $\bm{x}_t^d$ can be lower bounded for every $\epsilon\in\mathbb{R}$, $\bm{x}\in\mathbb{X}$, and sampling steps $i,j,t\in\mathbb{N}$ as the set is non-empty. Moreover, the probability of the control sequence sample $\hat{\bm{u}}_{[0:H]}(\bm{x}_t^d)$ being the global optimum $\bm{u}^*_{[0:H]}(\bm{x}_t^d)$ is positive by assumption. Thus, the probability of at least one sample in the set $\{(\bm{x}',\bm{u}_{[0:H]}): ||\bm{x}-\bm{x}'||\leq \epsilon, \bm{u}_{[0:H]}=\bm{u}^*_{[0:H]}(\bm{x}')\}$ can be bounded by a binomial distribution, which immediately implies that we can guarantee a sample in $\{\bm{x}': ||\bm{x}-\bm{x}'||\leq \epsilon\}$ with arbitrarily high probability by choosing sufficiently large $N_p+N_T+N_s$. 
\end{proof}
\else
\vspace{-0.45cm}
\fi

The assumption of positive probabilities $P(\bm{u}_{[0:H]}^*(\bm{x}))$ is necessary to ensure that global optimum can be found at all, but it does not pose a severe restriction in practice. In fact, this requirement can be easily satisfied using an initial guess distribution $\phi$ with positive support on a set $\mathbb{U}^H\supset\mathcal{U}^H$. Since the Gaussian perturbations of states in \cref{alg:data_generation} additionally induce a state distribution with support on $\mathbb{X}$, it immediately follows that \cref{alg:data_generation} is guaranteed to provide a suitable coverage of the globally optimal control sequences $\bm{u}^*_{[0:H]}(\bm{x}_t)$ when generating sufficiently large data sets $\mathcal{D}.$ Thereby, it increases the robustness against compounding errors without the necessity to train multiple policies as in on-policy imitation learning \citep{pmlr-v15-ross11a}. This is a crucial difference as the training of diffusion models exhibits a considerably higher computational complexity compared to shallow neural networks, which are the main focus of on-policy imitation learning.

\begin{algorithm}[t]
    \caption{Optimizer-based Data Generation}
    \begin{algorithmic}[1]
        \label{alg:data_generation}
        \REQUIRE{NMPC solver $\mathbb{S}$, perturbation variance $\bm{\sigma}$, initial guess distribution $\phi$, initial state distribution $\chi$, cost function $J$, sample numbers $N_s, N_T, N_p\in\mathbb{N}$}
        \FOR{$j=0,1,\cdots,N_s$}
            \STATE{Sample initial state $\bm{x}_0 \sim \chi$}
            \FOR{$t = 0,1,\cdots,N_T$}
                \FOR{$i=1,\ldots,N_p$}
                \STATE{Generate perturbed state $\bm{x}^d_t \gets \bm{x}_t + \bm{\epsilon}$, $\bm{\epsilon}\sim\mathcal{N}(\bm{0},\bm{\sigma}(t))$}
                \STATE{Sample initial guess  $\tilde{\bm{u}}_{[0:H]} \sim \phi$}
                \STATE{Execute local optimizer $\hat{\bm{u}}_{[0:H]}(\bm{x}_t^d) \gets \mathbb{S}(J(\bm{x}_t^d,\cdot),\tilde{\bm{u}}_{[0:H]})$}
                \STATE{Augment data set $\mathcal{D}\gets\mathcal{D}\cup \{\bm{x}_t^d,\hat{\bm{u}}_{[0:H]}(\bm{x}_t^d)\}$}
                \ENDFOR
                \STATE{Apply $\hat{\bm{u}}_0(\bm{x}_t)$ to system \eqref{eq:system}:  $ \bm{x}_{t+1} \gets \bm{f}(\bm{x}_t^d, \hat{\bm{u}}_0(\bm{x}_t^d))$}
            \ENDFOR
        \ENDFOR
    \end{algorithmic}
\end{algorithm}


\vspace{-0.2cm}
\subsection{Diffusion Model Training for Approximating NMPC} \label{sec:train}
\vspace{-0.1cm}

The conditional distribution $\rho(\cdot|\bm{x}_t)$, which is used to generate the data set $\mathcal{D}$ in \cref{sec:data_generation}, is potentially multi-modal due to the possible existence of multiple optima in the NMPC problem \eqref{eq:MPC_problem}. To learn a representation of this distribution, we leverage diffusion models due to their recent success in learning high-dimensional distributions from sample data \citep{pmlr-v37-sohl-dickstein15, NEURIPS2020_4c5bcfec}. The core idea behind diffusion models is an iterative denoising procedure, which is learned through an iterative noising process. 
Starting from the noiseless control sequence sample $\hat{\bm{u}}_{[0:H]}(\bm{x}_t)$ collected by \cref{alg:data_generation}, the noising process is based on iteratively adding Gaussian noise. The goal of this noising process is to perturb the distribution of optimal control sequences to Gaussian noise in the limit of infinite noising steps. More formally, this forward process, also known as the diffusion process, can be described by   
\begin{equation} \label{eq:diff_forward_eq}
    q(\bm{u}^{k}|\bm{u}^{k-1}):=\mathcal{N}(\bm{u}^{k};\sqrt{1-\beta_k}\bm{u}^{k-1}, \beta_k\mathbf{I}), \quad 1 \leq k \leq K \in \mathbb{N}, 
    \quad \bm{u}^0=\hat{\bm{u}}_{[0:H]}(\bm{x}_t),
\end{equation}
where $0 < \beta_1, \cdots, \beta_K < 1$ are scheduled variances. In terms of noise variables, this process is equivalently described using coefficients $\alpha_k := \prod_{j=1}^{k}(1-\beta_j)$, which results in
    \begin{equation} \label{eq:diff_forward_eq_arb_step}
        \bm{u}^{k} = \sqrt{\alpha_k}\bm{u}^0 + \sqrt{(1-\alpha_k)}\bm{\epsilon}, \quad \bm{\epsilon} \sim \mathcal{N}(\bm{0},\mathbf{I}).
    \end{equation}

Due to this forward process, the distribution of optimal control sequences can be described through a reverse process
    \vspace{-0.2cm}
    \begin{equation} \label{eq:diff_reverse}
            p_{\bm{\theta}}(\hat{\bm{u}}_{[0:H]}(\bm{x}_t))= \int p_{\bm{\theta}}(\bm{u}^{0:K})d\bm{u}^{1:K}= \int p(\bm{u}^K) \prod_{k=1}^K p_{\bm{\theta}}(\bm{u}^{k-1}|\bm{u}^{k})d\bm{u}^{1:K}
    \end{equation}
where the joint distribution $p_{\bm{\theta}}(\bm{u}^{0:K})$ is defined as a Markov chain with Gaussian transitions 
\begin{align}\label{eq:diff_reverse_conditional}
    p_{\bm{\theta}}(\bm{u}^{k-1}|\bm{u}^{k}):=\mathcal{N}(\bm{u}^{k-1};\bm{\mu}_{\bm{\theta}}({\bm{u}^k,k}),\bm{\Sigma}_k)
\end{align}
starting from a standard normal distribution $p(\bm{u}^K)$. In this reverse process, the mean functions $\bm{\mu}_{\bm{\theta}}$ of the conditional distributions $p_{\bm{\theta}}(\bm{u}^{k-1}|\bm{u}^k)$ can be represented by a neural network with parameters $\bm{\theta}$, which are optimized to represent the data distribution $\rho(\cdot|\bm{x}_t)$. When parameterized via $\bm{\mu}_{\bm{\theta}}(\bm{u}^k,k)=a_k \bm{u}^k+b_k \bm{\epsilon}_{\bm{\theta}}(\bm{u}^k,k)$ with $a_k=\nicefrac{1}{\sqrt{1-\beta_k}}$ and $b_k=\nicefrac{-\beta_k}{(\sqrt{1-\beta_k}\sqrt{1-\alpha_k})}$, training the parameters $\bm{\theta}$ can be effectively performed by minimizing the loss $\mathbb{E}_{k \in [1,K],\bm{\epsilon} \thicksim \mathcal{N}(\bm{0},\mathbf{I})} [||\bm{\epsilon}-\bm{\epsilon}_{\bm{\theta}}(\bm{u}^k,k) ||^2]$, with the predicted noise $\bm{\epsilon}_{\bm{\theta}}(\bm{u}^k,k)$. Hence, we essentially train a neural network $\bm{\epsilon}_{\bm{\theta}}$ using the value $\bm{u}^k$ defined in the forward process \eqref{eq:diff_forward_eq_arb_step} to obtain the mean $\bm{\mu}_{\bm{\theta}}$ in the conditional distributions \eqref{eq:diff_reverse_conditional}.

While diffusion models have been originally developed to learn a fixed distribution, it is straightforward to extend them to learning state-dependent distributions as needed for control. For this purpose, we apply the classifier-free diffusion guidance \citep{ho2022classifier} to learn a conditional diffusion model. In this method, the noise predictor integrates both conditional information $\bm{\epsilon}_{\bm{\theta}}(\bm{u}^k,k,\bm{x}_t)$ and unconditional information $\bm{\epsilon}_{\bm{\theta}}(\bm{u}^k,k,\varnothing)$ during training, where $\varnothing$ is a null token for the condition $\bm{x}_t$. Therefore, it employs a training loss
    \begin{equation} \label{eq:MPDC_loss}
        \mathcal{L} = \mathbb{E}_{k \in [1,K],\bm{\epsilon} \thicksim \mathcal{N}(\bm{0},\mathbf{I}), \bm{x}_t \thicksim \mathcal{D}, b \sim \text{Bernoulli}(p)} [||\bm{\epsilon}-\bm{\epsilon}_{\bm{\theta}}(\bm{u}^k,k,(1-b)\cdot \bm{x}_t + b \cdot \varnothing) ||^2_2],
    \end{equation}
with a probability $p \in [0, 1]$ for the unconditioned information. Since diffusion models are capable of capturing arbitrary data distributions given sufficiently flexible model architectures $p_{\bm{\theta}}$ \citep{pmlr-v37-sohl-dickstein15}, this approach enables us to learn an arbitrarily accurate approximation $p_{\bm{\theta}}(\cdot|\bm{x}_t)$ of the distribution $\rho(\cdot|\bm{x}_t)$ over optimal control sequences $\hat{\bm{u}}_{[0:H]}(\bm{x}_t)$ given a sufficiently large data set $\mathcal{D}$. Hence, for every $\bm{x}\in\mathbb{X}$ and $\bm{u}_{[0:H]}\in\mathbb{U}^H$, we can achieve 
\begin{align}\label{eq:learn_error}
    \left|\int_{\bm{\upsilon}\in\mathbb{B}_{\varepsilon}(\bm{u}_{[0:H]})} \rho(\bm{\upsilon}|\bm{x}_t)-p_{\bm{\theta}}(\bm{\upsilon}|\bm{x}_t) \mathrm{d}\bm{\upsilon}\right|\leq \tilde{\delta},
\end{align}
where $\mathbb{B}_{\varepsilon}(\bm{u}_{[0:H]})$ is a ball with radius $\varepsilon$ and center $\bm{u}_{[0:H]}$, and $\tilde{\delta} \in (0,1]$ is a probability depending on $\bar{\delta}$ in Theorem~\ref{th:data}.\looseness=-1


\vspace{-0.1cm}
\subsection{Toward Near-Globally Optimal NMPC using Diffusion Models} 
\label{sec:inference}
\vspace{-0.1cm}

The diffusion model provides us with a representation $p_{\bm{\theta}}$ of the probability distribution of optimal control sequences $\rho$, which does not involve the application of the numerical solver $\mathbb{S}$. Sampling from $p_{\bm{\theta}}(\cdot|\bm{x}_t)$ is particularly straighforward: After sampling an initial control sequence $\bm{u}^K\sim\mathcal{N}(0,\bm{I})$, we iteratively sample $\bm{u}^{k-1}\sim p_{\bm{\theta}}(\cdot|\bm{u}^k)$ from the conditional distributions \eqref{eq:diff_reverse_conditional} until we obtain $\hat{\bm{u}}_{[0:H]}^{m}=\bm{u}^0$. Therefore, we can sample $\hat{\bm{u}}_{[0:H]}^{m}\sim p_{\bm{\theta}}(\cdot|\bm{x}_t)$ through an iterative procedure.\looseness=-1 

\begin{algorithm}[t]
	\caption{ NMPC via Diffusion Models}
	\begin{algorithmic}[1]
		\label{alg:mpc_diff}
		\REQUIRE{diffusion model $p_{\bm{\theta}}(\cdot|\cdot)$, number of optimization samples $M$}
        \WHILE{$\TRUE$}
        \STATE{Measure system state $\bm{x}_t$}
        \FOR{$m=1,\cdots,M$}
            \STATE{Sample control sequence $\hat{\bm{u}}^m_{[0:H]} \sim p_{\bm{\theta}}(\cdot|\bm{x}_t)$} 
            \STATE{Compute cost value $v^m \gets J(\bm{x}_t, \hat{\bm{u}}^m_{[0:H]})$}
        \ENDFOR
        \STATE{Obtain control sequence $m^* \gets \argmin_{m=1,\ldots,M} v^m$} 
        \STATE Apply control input $\bm{x}_{t+1}\gets \bm{f}(\bm{x}_t,\hat{\bm{u}}^{m^*}_0)$, $t\gets t+1$
        \ENDWHILE
	\end{algorithmic}
\end{algorithm}

Since we can ensure that the diffusion model exhibits a low approximation error, i.e., $p_{\bm{\theta}}(\cdot|\bm{x})\approx \rho(\cdot|\bm{x})$, this possibility for approximately sampling from the distribution of optimal control sequences $\hat{\bm{u}}_{[0:H]}(\bm{x}_t)$ provides us with a computationally efficient approach toward near-globally optimal NMPC as outlined in \cref{alg:mpc_diff}. Instead of executing a numerical optimizer, we perform a random shooting approach by sampling multiple control sequences $\hat{\bm{u}}_{[0:H]}^{m}\sim p_{\bm{\theta}}(\cdot|\bm{x}_t)$, $m=1,\ldots,M$ with $M\in\mathbb{N}$. For each sample $\hat{\bm{u}}_{[0:H]}^{m}$, we compute the corresponding cost $v^m=J(\bm{x}_t,\hat{\bm{u}}_{[0:H]}^{m})$, such that a simple comparison allows to select the best control sequence $\bm{u}^{m^*}_{[0:H]}$, where $m^*= \argmin_{m=1,\ldots,M} v^m$. The result $\bm{u}^{m^*}_{[0:H]}$ of this sample, score, and rank approach is a random variable, whose distance to the globally optimal control sequence $\bm{u}^*_{[0:H]}(\bm{x}_t)$ can be probabilistically bounded to ensure near-global optimality as defined in \eqref{eq:control_goal}. 
\vspace{-0.1cm}
\begin{theorem}\label{th:opt}
    Consider a diffusion model satisfying \eqref{eq:learn_error} with parameters $\tilde{\delta},\varepsilon$ for all $\bm{x}\in\mathbb{X}$. If 
    $\tilde{\delta}<p_{\mathbb{B}}=\int_{\bm{\upsilon}\in\mathbb{B}_{\varepsilon}(\bm{u}_{[0:H]}(\bm{x}_t))} \rho(\bm{\upsilon}|\bm{x}_t) \mathrm{d}\bm{\upsilon}$
    for all $\bm{x}\in\mathbb{X}$ and balls $\mathbb{B}_{\varepsilon}(\bm{u}_{[0:H]}(\bm{x}_t))$ with radius $\varepsilon$ and center $\bm{u}_{[0:H]}(\bm{x}_t)$, then, the policy defined by Algorithm~\ref{alg:mpc_diff} is probabilistically near-globally optimal, i.e., it satisfies \eqref{eq:control_goal}, with constants $\varepsilon$ and $\delta=(1-p_{\mathbb{B}}+\tilde{\delta})^M$.
\end{theorem}
\if\isarxiv1
\begin{proof}
    The probability that a single sample from the diffusion model within a distance of $\varepsilon$ from the global optimum $\bm{u}^*_{[0:H]}(\bm{x}_t)$ is given by
    \begin{align}
        P(||\hat{\bm{u}}_{[0:H]}^m- \bm{u}^*_{[0:H]}(\bm{x}_t)||_{2}\leq \varepsilon)= \int_{\bm{\upsilon}\in\mathbb{B}_{\varepsilon}(\bm{u}_{[0:H]})} p_{\bm{\theta}}(\bm{\upsilon}|\bm{x}_t) \mathrm{d}\bm{\upsilon}.
    \end{align}
    Using the error bound \eqref{eq:learn_error}, we can bound this probability by
    \begin{align}
    \int_{\bm{\upsilon}\in\mathbb{B}_{\varepsilon}(\bm{u}_{[0:H]})} p_{\bm{\theta}}(\bm{\upsilon}|\bm{x}_t) \mathrm{d}\bm{\upsilon}&= 
    \int_{\bm{\upsilon}\in\mathbb{B}_{\varepsilon}(\bm{u}_{[0:H]})} \rho(\bm{\upsilon}|\bm{x}_t) \mathrm{d}\bm{\upsilon}+\int_{\bm{\upsilon}\in\mathbb{B}_{\varepsilon}(\bm{u}_{[0:H]})} p_{\bm{\theta}}(\bm{\upsilon}|\bm{x}_t) -\rho(\bm{\upsilon}|\bm{x}_t)\mathrm{d}\bm{\upsilon} \nonumber\\
    &\geq \int_{\bm{\upsilon}\in\mathbb{B}_{\varepsilon}(\bm{u}_{[0:H]})} \rho(\bm{\upsilon}|\bm{x}_t) \mathrm{d}\bm{\upsilon}-\left|\int_{\bm{\upsilon}\in\mathbb{B}_{\varepsilon}(\bm{u}_{[0:H]})} \rho(\bm{\upsilon}|\bm{x}_t)-p_{\bm{\theta}}(\bm{\upsilon}|\bm{x}_t) \mathrm{d}\bm{\upsilon}\right| \nonumber\\
    &\geq p_{\mathbb{B}}-\tilde{\delta}\nonumber\\
    &>0,
\end{align}
where the last inequality holds due to the assumed lower bound on $\int_{\bm{\upsilon}\in\mathbb{B}_{\varepsilon}(\bm{u}_{[0:H]}(\bm{x}_t))} \rho(\bm{\upsilon}|\bm{x}_t) \mathrm{d}\bm{\upsilon}> \tilde{\delta}$. Since all sampled control sequences are independent, we additionally have
\begin{align}
    P(||\hat{\bm{u}}_{[0:H]}^m- \bm{u}^*_{[0:H]}(\bm{x}_t)||_{2}\leq \varepsilon) &= 1-P(||\hat{\bm{u}}_{[0:H]}^m- \bm{u}^*_{[0:H]}(\bm{x}_t)||_{2}> \varepsilon, \forall m=1,\ldots,M)\nonumber\\ 
    &\geq 1-(1-p_{\mathbb{B}}+\tilde{\delta})^M.
\end{align}
Thus, probabilistic, near-global optimality is guaranteed with constants $\varepsilon$ and $\delta=(1-p_{\mathbb{B}}+\tilde{\delta})^M$.
\end{proof}
\else
\vspace{-0.1cm}
\fi
This result guarantees that we can find an arbitrarily close approximation of the globally optimal policy with high probability, with the particular quality of the approximation accuracy depending on the number of generated training samples via Theorem \ref{th:data} and the expressivity of the employed diffusion model. This guarantee straightforwardly follows from a positive probability that a sample falls within an $\varepsilon$-ball around the global optimum for sufficiently high model accuracy, i.e., $\tilde{\delta}<p_{\mathbb{B}}$.

Since the samples are independent, we can multiply these probabilities to bound the probability $\delta$ that none of the samples is close to $\bm{u}^*_{[0:H]}(\bm{x}_t)$, which yields $\delta=(1-p_{\mathbb{B}}+\tilde{\delta})^M$. 
Hence, $\delta$ can be made arbitrarily small by choosing a sufficiently large $M$, such that computational online complexity and accuracy can be effectively balanced.

This trade-off is particularly beneficial since sampling from the diffusion model is highly parallelizable. While we can also make use of parallelization for global optimization using multiple initializations of local optimizers, existing fast solver implementations such as ACADOS \citep{Verschueren2021} and FATROP \citep{vanroye2023fatrop} are restricted to CPUs. This severely limits the number of OCPs that can be solved in parallel in practice. In contrast, diffusion models are designed for an evaluation on GPUs whose success is based on parallelization, such that the degree of parallelization can be orders of magnitude higher. Thereby, our diffusion model-based approach allows the generation of larger numbers of samples, which implies control sequences closer to the global optimum due to Theorem~\ref{th:opt}.

\if\isarxiv1
\begin{remark}
    In offline RL, it has been proposed to condition the diffusion model additionally on the normalized cost, such that the optimal control input can directly be sampled from the diffusion model  \citep{ajay2023is}. Thereby, the additional random shooting optimization employed in our approach is avoided. However, this requires a normalization of the cost values, which necessitates global optimization during data generation for state-dependent normalization, or similar cost within the entire state space $\mathbb{X}$ for global normalization. While the former is undesirable due to the complexity of global optimization, the latter is generally not true with commonly used NMPC cost functions. Therefore, conditioning on the normalized cost values is not applicable in our setting.
\end{remark}
\fi

\vspace{-0.1cm}
\section{Numerical Evaluation}\label{numerical_result}
\vspace{-0.1cm}

\setlength{\textfloatsep}{7pt}

In this section, we evaluate our proposed framework to answer the following questions:
(1) Can our approach learn an effective representation of the optima of OCP solved for NMPC? (2) Does it provide computational and performance advantages compared to baselines? 

\vspace{-0.1cm}
\subsection{Simulation Setting} \label{sec:sim_setting}
\vspace{-0.1cm}

We apply our approach to the following dynamic systems: a) cart-pole, b) pendubot, and c) double cart-pole in our evaluation. The initial condition for each dynamics is such that the pole rests in a neighborhood of the downward position. The control task is to perform a swing-up to the unstable upward position. For this purpose, we define quadratic cost functions for each system, which are summed up over horizons of 64 (a), 256 (b), and 128 (c) time steps. Initial states are sampled from a uniform distribution defined over a neighborhood of the stable downward position. The control task for these systems is well-suited for the investigation of our approach since we know that with the left and right swing up, generally, at least two locally optimal solutions of the OCP exist. As illustrated on the left side of \cref{fig:ablation+multimod} for the initial state distribution of the cart-pole system, this leads to a multi-modal distribution of the NMPC solutions in general.

For all the dynamic systems, \cref{alg:data_generation} is used to generate the data set for the representation of the multi-modal distribution with the sampling time $0.01s$. The initial guesses are sampled from a uniform distribution $\phi$, and the range of $\phi$ is adjusted to $[-\bar{u}_{t-1}, \bar{u}_{t-1}], t=1,2,\cdots,N_T$, where $\bar{u}_0 \in \mathbb{U}$. The numerical optimization is executed using CasADi \citep{andersson2019casadi} with an interior-point optimizer (IPOPT) \citep{ipopt}. The diffusion model is trained using the loss \eqref{eq:MPDC_loss} with probability $p=0.25$ for the unconditioned information.
\if\isarxiv1
More details regarding our simulation setup can be found in the appendix.
\else
More details regarding our simulation setup can be found in the extended version of this paper~\footref{fn:extension}.
\fi

\begin{figure}[t]
    \centering
            \begin{tikzpicture}
            \def\file{figure/uDistribution.txt}
            \begin{axis}[
            height=3.4cm, width = \textwidth/2.2,
            axis y line*=right,
            axis x line=none,
            ymin=-6500, ymax=9900,
            xmin = -2, xmax = 52,
            ylabel={\sffamily control input },
            ]
            \addplot[orange!80!white, name path=f1r,thick] table [x = f1r, y = xf1]{\file};
            \addplot[orange!80!white, name path=f1l,thick] table [x = f1l, y = xf1]{\file};
            \addplot[orange!50!white,opacity=0.6, forget plot] fill between[of=f1r and f1l];	

            \addplot[orange!80!white, name path=f2r,thick] table [x = f2r, y = xf2]{\file};
            \addplot[orange!80!white, name path=f2l,thick] table [x = f2l, y = xf2]{\file};
            \addplot[orange!50!white,opacity=0.6, forget plot] fill between[of=f2r and f2l];	

            \addplot[orange!80!white, name path=f3r,thick] table [x = f3r, y = xf3]{\file};
            \addplot[orange!80!white, name path=f3l,thick] table [x = f3l, y = xf3]{\file};
            \addplot[orange!50!white,opacity=0.6, forget plot] fill between[of=f3r and f3l];	

            \addplot[orange!80!white, name path=f4r,thick] table [x = f4r, y = xf4]{\file};
            \addplot[orange!80!white, name path=f4l,thick] table [x = f4l, y = xf4]{\file};
            \addplot[orange!50!white,opacity=0.6, forget plot] fill between[of=f4r and f4l];	

            \addplot[orange!80!white, name path=f5r,thick] table [x = f5r, y = xf5]{\file};
            \addplot[orange!80!white, name path=f5l,thick] table [x = f5l, y = xf5]{\file};
            \addplot[orange!50!white,opacity=0.6, forget plot] fill between[of=f5r and f5l];	

            \addplot[orange!80!white, name path=f6r,thick] table [x = f6r, y = xf6]{\file};
            \addplot[orange!80!white, name path=f6l,thick] table [x = f6l, y = xf6]{\file};
            \addplot[orange!50!white,opacity=0.6, forget plot] fill between[of=f6r and f6l];	
            \end{axis}
            \def\file{figure/stateMultiModality.txt}
            \begin{axis}[ylabel={\sffamily angle $\theta$},
            xlabel={\sffamily \# time step $t$},
            xlabel shift = -0.15cm,
                xmin=-2, ymin = -0.5, xmax = 52, ymax =9.5, 
            height =3.4cm, width=\textwidth/2.2,
            legend pos = north west, legend columns=3, legend entries = {\sffamily left swing, \sffamily right swing, \sffamily  control distr.},
            name=plot1, ylabel near ticks, ytick pos=left
            ]

            \addplot[name path=lower1, fill=none, draw=none, forget plot]	table[x = step, y  = positive_mode_1_max]{\file};
            \addplot[name path=upper1, fill=none, draw=none, forget plot]	table[x = step, y = positive_mode_1_min]{\file};
            \addplot[blue!40,opacity=0.4, forget plot] fill between[of=lower1 and upper1];	

            \addplot[name path=lower1, fill=none, draw=none, forget plot]	table[x = step, y  = negative_mode_1_max]{\file};
            \addplot[name path=upper1, fill=none, draw=none, forget plot]	table[x = step, y = negative_mode_1_min]{\file};
            \addplot[green!40,opacity=0.4, forget plot] fill between[of=lower1 and upper1];
            
            \addplot[blue!90!black, very thick] table [x = step, y = positive_mode_1_mean]{\file};
            \addplot[green!70!black, very thick] table [x = step, y = negative_mode_1_mean]{\file};

            \addlegendimage{area legend, fill=orange!50!white, draw=orange!80!white};

            \end{axis}
            \end{tikzpicture}
            \def\file{figure/PtgMultiModality.txt}
            \begin{tikzpicture}
            \begin{axis}[ylabel={\sffamily multimodality [\%]},
            xlabel={\sffamily \# time step $t$},
            xlabel shift = -0.15cm,
            xmin=0.0, ymin = 0, xmax = 50, ymax =100, 
            height =3.4cm, width=\textwidth/2.2,
            legend pos = north east, legend columns=1, legend entries = {\sffamily our method, \sffamily global optim.},
            name=plot1, ylabel near ticks
         ]
            \addplot[blue!90!black, very thick] table [x = step, y = OurMethod]{\file};
            \addplot[orange, very thick] table [x = step, y = MPCMG]{\file};
            
            \end{axis}
            \end{tikzpicture}

    \vspace{-0.7cm}
\caption{Left: Evolution of the pole angle $\theta$ resulting from two kinds of optimal control sequences, whose multimodal distribution is illustrated for different time steps. Right: Comparison of the multimodality of samples from our trained diffusion model and samples employed by a global optimizer for different time steps along roll-out trajectories. }
\label{fig:ablation+multimod}
\end{figure}

We compare our diffusion model approach for
NMPC to the following baseline methods:
\vspace{-0.2cm}
\begin{itemize}
    \item $\textbf{Locally Optimal Model Predictive Control}$: We use CasADI \citep{andersson2019casadi} with IPOPT \citep{ipopt} as a baseline of a common local optimization-based NMPC method. The results from previous time steps are used to warm-start the optimization.\looseness=-1
    \vspace{-0.2cm}
    \item $\textbf{Globally Optimal Model Predictive Control}$: As a global optimization NMPC approach, we use CasADI \citep{andersson2019casadi} with mulitple random initializations. 
    Thereby, this approach also relies on a sample, rank, and score approach, but requires execution of the local optimizer for sampling.
    \vspace{-0.2cm}
    \item $\textbf{Neural Network Approximation of Model Predictive Control (NN)}$: We use the data set generated by \cref{alg:data_generation} to train a neural network model of MPC in a supervised fashion as an example of behavior cloning suggested, e.g., by \cite{8371312}. The neural network has $3$ hidden layers, each comprised of $50$ neurons with $\tanh$ activation functions. 
    \vspace{-0.2cm}
    \item $\textbf{Neural Network Approximation of Globally-
    Optimal Model Predictive Control (NN${}^*$)}$: \linebreak
    We sample multiple random initial guesses to create a data set of approximately globally optimal control sequences, which are used to train a neural network approximation of globally optimal NMPC in a supervised fashion. To ensure a fair comparison, we keep the overall number of solved OCPs constant by setting reduced values of $N_s, N_T$, and $N_p$ in \cref{alg:data_generation}.
    \vspace{-0.2cm}
\end{itemize}
All the algorithms are implemented in Python, and the simulations are conducted on a machine with an AMD EPYC 7542 32 cores processor and NVIDIA Tesla P100. Note that we refrain from using optimized implementations of NMPC solvers such as {CASADOS} \citep{Verschueren2021} or FATROP \citep{vanroye2023fatrop} since our diffusion model implementation is also not exploiting techniques for accelerated code execution.

\vspace{-0.1cm}
\subsection{Result in Simulation}
\vspace{-0.1cm}

\begin{figure}
    \centering
    \begin{tikzpicture}
	\begin{axis}[
        name = plot1,
        width = \textwidth/2.5, height=3.0cm,
		boxplot/draw direction = y,
		enlarge y limits,
		xtick = {1, 2, 3, 4, 5},
		xticklabel style = {align=center, font=\small, rotate=25},
		xticklabels = {},
		xtick style = {draw=none}, 
		ylabel = {\textsf{norm. cost}},
        ylabel near ticks,
        title={\sffamily a) cart-pole}
	]  
        \addplot+[
        fill = blue!30, draw=blue, 
        boxplot prepared={
        median=0.075969,
        upper quartile=0.12213,
        lower quartile=0.020962,
        upper whisker=0.211803,
        lower whisker=0.000000
        }] coordinates {};

        \addplot+[
        fill = orange!30, draw=orange, 
        boxplot prepared={
        median=0.812931,
        upper quartile=0.926821,
        lower quartile=0.692513,
        upper whisker=1,
        lower whisker=0.518567
        }] coordinates {};

        \addplot+[
        fill = green!20, draw=green!50!black, 
        boxplot prepared={
        median=0.468492,
        upper quartile=0.645026,
        lower quartile=0.362787,
        upper whisker=0.923574,
        lower whisker=0.277712
        }] coordinates {};

        \addplot+[
        fill = yellow!30, draw=yellow!75!black, 
        boxplot prepared={
        median=0.544508,
        upper quartile=0.710989,
        lower quartile=0.317097,
        upper whisker=0.816997,
        lower whisker=0.131544
        }] coordinates {};

        \addplot+[
        fill = purple!30, draw=purple, 
        boxplot prepared={
        median=0.069424,
        upper quartile=0.103138,
        lower quartile=0.019277,
        upper whisker=0.218698,
        lower whisker=0.005155
        }] coordinates {};

	\end{axis}
    \begin{axis}[
        name = plot2,
        width = \textwidth/2.5, height=3.0cm,
		boxplot/draw direction = y,
		enlarge y limits,
		xtick = {1, 2, 3, 4, 5},
		xticklabel style = {align=center, font=\small, rotate=25},
		xticklabels = {},
		xtick style = {draw=none}, 
        ylabel near ticks,
        yticklabels  = {},
        at=(plot1.east), anchor=east, xshift=4.7cm,
        title={\sffamily b) pendubot}
	]  
        \addplot+[
        fill = blue!30, draw=blue, 
        boxplot prepared={
        median=0.109129,
        upper quartile=0.148604,
        lower quartile=0.065449,
        upper whisker=0.211339,
        lower whisker=0.055396
        }] coordinates {};

        \addplot+[
        fill = orange!30, draw=orange, 
        boxplot prepared={
        median=0.895271,
        upper quartile=0.950041,
        lower quartile=0.809200,
        upper whisker=1,
        lower whisker=0.661222
        }] coordinates {};

        \addplot+[
        fill = green!20, draw=green!50!black, 
        boxplot prepared={
        median=0.285931,
        upper quartile=0.337658,
        lower quartile=0.253387,
        upper whisker=0.432233,
        lower whisker=0.238265
        }] coordinates {};

        \addplot+[
        fill = yellow!30, draw=yellow!75!black, 
        boxplot prepared={
        median=0.416412,
        upper quartile=0.815209,
        lower quartile=0.166112,
        upper whisker=0.818787,
        lower whisker=0.000243
        }] coordinates {};

        \addplot+[
        fill = purple!30, draw=purple, 
        boxplot prepared={
        median=0.046578,
        upper quartile=0.123182,
        lower quartile=0.035955,
        upper whisker=0.131749,
        lower whisker=0.000000
        }] coordinates {};
	\end{axis}
    \begin{axis}[
        name = plot3,
        width = \textwidth/2.5, height=3.0cm,
		boxplot/draw direction = y,
		enlarge y limits,
		xtick = {1, 2, 3, 4, 5},
		xticklabel style = {align=center, font=\small, rotate=25},
		xticklabels = {},
		xtick style = {draw=none}, 
        ylabel near ticks,
        yticklabels  = {},
        at=(plot2.east), anchor=east, xshift=4.7cm,
        title={\sffamily c) double cart-pole}
	]  
        \addplot+[
        fill = blue!30, draw=blue, 
        boxplot prepared={
        median=0.036858,
        upper quartile=0.065771,
        lower quartile=0.019421,
        upper whisker=0.124783,
        lower whisker=0.0
        }] coordinates {};

        \addplot+[
        fill = orange!30, draw=orange, 
        boxplot prepared={
        median=0.273915,
        upper quartile=0.281359,
        lower quartile=0.266056,
        upper whisker=0.290548,
        lower whisker=0.264359
        }] coordinates {};

        \addplot+[
        fill = green!20, draw=green!50!black, 
        boxplot prepared={
        median=0.672566,
        upper quartile=0.672566,
        lower quartile=0.672566,
        upper whisker=0.672566,
        lower whisker=0.672566
        }] coordinates {};

        \addplot+[
        fill = yellow!30, draw=yellow!75!black, 
        boxplot prepared={
        median=0.280256,
        upper quartile=0.532490,
        lower quartile=0.212921,
        upper whisker=1,
        lower whisker=0.020561
        }] coordinates {};

        \addplot+[
        fill = purple!30, draw=purple, 
        boxplot prepared={
        median=0.034701,
        upper quartile=0.036609,
        lower quartile=0.023017,
        upper whisker=0.037384,
        lower whisker=0.020561
        }] coordinates {};
	\end{axis}
 
	\begin{axis}[
        name = plot4,
        width = \textwidth/2.5, height=3.0cm,
		boxplot/draw direction = y,
		enlarge y limits,
		xtick = {1, 2, 3, 4, 5},
		xticklabel style = {align=center, font=\small, rotate=20},
		xticklabels = {\sffamily\tiny diffusion, \sffamily\tiny NN, \sffamily\tiny NN${}^*$, \sffamily\tiny local optim., \sffamily\tiny global optim.},
		xtick style = {draw=none}, 
		ylabel = {\textsf{comp. time [s]}~~~},
        ylabel near ticks,
        at=(plot1.east), anchor=east, yshift=-1.6cm,
        ymax = 8.5, ymin=0.32
	]  
        \addplot+[
        fill = blue!30, draw=blue, 
        boxplot prepared={
        median=0.404,
        upper quartile=0.406,
        lower quartile=0.403,
        upper whisker=0.420,
        lower whisker=0.401
        }] coordinates {};

        \addplot+[
        fill = orange!30, draw=orange, 
        boxplot prepared={
        median=0.000276,
        upper quartile=0.000278,
        lower quartile=0.000276,
        upper whisker=0.000335,
        lower whisker=0.000274
        }] coordinates {};

        \addplot+[
        fill = green!20, draw=green!50!black, 
        boxplot prepared={
        median=0.000276,
        upper quartile=0.000278,
        lower quartile=0.000275,
        upper whisker=0.000271,
        lower whisker=0.000344
        }] coordinates {};

        \addplot+[
        fill = yellow!30, draw=yellow!75!black, 
        boxplot prepared={
        median=0.316,
        upper quartile=0.340,
        lower quartile=0.301,
        upper whisker=0.449,
        lower whisker=0.251
        }] coordinates {};

        \addplot+[
        fill = purple!30, draw=purple, 
        boxplot prepared={
        median=1.525,
        upper quartile=1.682,
        lower quartile=1.435,
        upper whisker=2.034,
        lower whisker=1.347
        }] coordinates {};
	\end{axis}
    \begin{axis}[
        name = plot5,
        width = \textwidth/2.5, height=3.0cm,
		boxplot/draw direction = y,
		enlarge y limits,
		xtick = {1, 2, 3, 4, 5},
		xticklabel style = {align=center, font=\small, rotate=20},
		xticklabels = {\sffamily\tiny diffusion, \sffamily\tiny NN, \sffamily\tiny NN${}^*$, \sffamily\tiny local optim., \sffamily\tiny global optim.},
		xtick style = {draw=none}, 
        yticklabels = {},
        ylabel near ticks,
        at=(plot2.east), anchor=east, yshift=-1.6cm,
        ymax = 8.5,ymin=0.32
	]  
        \addplot+[
        fill = blue!30, draw=blue, 
        boxplot prepared={
        median=0.341,
        upper quartile=0.343,
        lower quartile=0.339,
        upper whisker=0.345,
        lower whisker=0.336
        }] coordinates {};

        \addplot+[
        fill = orange!30, draw=orange, 
        boxplot prepared={
        median=0.000360,
        upper quartile=0.000367,
        lower quartile=0.000346,
        upper whisker=0.000384,
        lower whisker=0.000339
        }] coordinates {};

        \addplot+[
        fill = green!20, draw=green!50!black, 
        boxplot prepared={
        median=0.000434,
        upper quartile=0.000438,
        lower quartile=0.000431,
        upper whisker=0.000446,
        lower whisker=0.000426
        }] coordinates {};

        \addplot+[
        fill = yellow!30, draw=yellow!75!black, 
        boxplot prepared={
        median=1.314,
        upper quartile=1.353,
        lower quartile=1.283,
        upper whisker=1.422,
        lower whisker=1.243
        }] coordinates {};

        \addplot+[
        fill = purple!30, draw=purple, 
        boxplot prepared={
        median=6.966,
        upper quartile=7.140,
        lower quartile=6.855,
        upper whisker=7.540,
        lower whisker=6.612
        }] coordinates {};
	\end{axis}
    \begin{axis}[
        name = plot6,
        width = \textwidth/2.5, height=3.0cm,
		boxplot/draw direction = y,
		enlarge y limits,
		xtick = {1, 2, 3, 4, 5},
		xticklabel style = {align=center, font=\small, rotate=20},
		xticklabels = {\sffamily\tiny diffusion, \sffamily\tiny NN, \sffamily\tiny NN${}^*$, \sffamily\tiny local optim., \sffamily\tiny global optim.},
		xtick style = {draw=none}, 
        yticklabels = {},
        ylabel near ticks,
        at=(plot3.east), anchor=east, yshift=-1.6cm, 
        ymax = 8.5, ymin=0.32
	]  
        \addplot+[
        fill = blue!30, draw=blue, 
        boxplot prepared={
        median=0.359,
        upper quartile=0.359,
        lower quartile=0.358,
        upper whisker=0.361,
        lower whisker=0.357
        }] coordinates {};

        \addplot+[
        fill = orange!30, draw=orange, 
        boxplot prepared={
        median=0.000394,
        upper quartile=0.000400,
        lower quartile=0.000391,
        upper whisker=0.000411,
        lower whisker=0.000389
        }] coordinates {};

        \addplot+[
        fill = green!20, draw=green!50!black, 
        boxplot prepared={
        median=0.000374,
        upper quartile=0.000378,
        lower quartile=0.000372,
        upper whisker=0.000382,
        lower whisker=0.000369
        }] coordinates {};

        \addplot+[
        fill = yellow!30, draw=yellow!75!black, 
        boxplot prepared={
        median=1.243,
        upper quartile=1.709,
        lower quartile=1.037,
        upper whisker=2.716,
        lower whisker=0.736
        }] coordinates {};

        \addplot+[
        fill = purple!30, draw=purple, 
        boxplot prepared={
        median=6.296,
        upper quartile=7.145,
        lower quartile=5.790,
        upper whisker=9.013,
        lower whisker=5.002
        }] coordinates {};
	\end{axis}
\end{tikzpicture}
\vspace{-0.7cm}
\caption{Top: Our diffusion model-based approach achieves similar cost as a global optimization approach for NMPC, but significantly outperforms direct neural network approximations of NMPC. Bottom: The required computation times for online optimization based on a diffusion model are significantly reduced compared to the direct global optimization approach and at most on par with the commonly employed local optimizer. 
}
\label{fig:performance}
\end{figure}

We first investigate question (1) regarding the capability of our proposed diffusion model approach for learning a multimodal representation of the optimal control sequences. To answer this question, we take the cart-pole system as an example and compare our approach to NMPC with multiple initial guesses as introduced in \cref{sec:sim_setting}. 
Starting from $100$ initial states, we compute the roll-out trajectories of both control laws for $50$ time steps. For each state $\bm{x}_t$, $M\!=\!20$ samples are drawn from $\rho(\cdot|\bm{x}_t)$ through numerical optimization or via the diffusion model. Therefore, we can measure the multi-modality of the distributions, i.e., if multiple optima have been found, by determining if the distance between any of the $20$ control sequences exceeds a threshold. 
\if\isarxiv1
We empirically found that $70$ is a suitable value for this threshold as it is sufficiently large to avoid false positives due to inaccurate optimization. 
\fi
The resulting percentages for finding multiple optima are illustrated on the right side of \cref{fig:ablation+multimod}, which clearly demonstrates that our diffusion model is capable of achieving the same diversity as obtained with randomized optimization in NMPC. Note that the loss of multi-modality at later time steps is due to the nature of the considered control problem: once a swing-up in one direction has been executed for a certain number of time steps, the local minimum corresponding to the other swing-up direction becomes extremely large, such that it eventually disappears. Therefore, our diffusion model approach for NMPC is well-suited to reflect the distribution of optimal control sequences for the cartpole system.\looseness=-1

Based on this insight, we compare our diffusion model approach for NMPC with baseline methods to investigate its advantages, thereby answering the question (2). For this, we sample initial states from $\chi$ randomly for each method and generate roll-outs for the swing-up control task. For each method, we determine the cost experienced during the closed-loop control roll-outs and measure the computation times needed for obtaining the control inputs. The results of this comparison are depicted in \cref{fig:performance}. It can be clearly seen that our diffusion model approach for NMPC achieves a lower cost than all baseline methods in this scenario except for the NMPC approach with multiple initial guesses, which performs similarly. Since this approach directly employs global optimization, it poses a lower bound on the achievable cost underlining the strength of our method. It is worth noting that neural network-based approximations of NMPC benefit from lower computational times due to fast forward inference, but struggle to capture the multimodal nature of the underlying control distributions and consequently exhibit inferior performance. Furthermore, approaches requiring globally optimized training targets suffer from limited data availability, negatively impacting generalization. While our diffusion model approach is a significant improvement over the neural network approximation of the globally optimal NMPC scheme for all the investigated system dynamics, its complexity is comparable to or lower than the locally optimal NMPC method. We emphasize that our diffusion model has not been optimized for fast inference, such that significantly lower complexity can be expected using improved sampling schemes \citep{song2020denoising,frans2024one}.\looseness=-1

While advantages of our diffusion model approach can be observed for the considered exemplary systems, they only exhibit a modest complexity. However, the computational benefits of the diffusion model become significantly more pronounced when a more challenging optimization landscape requires many more control sequence samples $M$ or when longer prediction horizons $H$ are considered, as illustrated on the left and center of Fig.~\ref{fig:inferenceNum_compare}, respectively. While the performance of our diffusion model approach to NMPC closely reflects the performance of the globally optimal MPC scheme for the cart-pole system over a range of parameters $M$ and $H$, the computation time barely changes. In contrast, a steep increase in the computation time is visible for the globally optimal MPC method due to the increasing number of optimizations and growing complexity of each optimization, respectively. Finally, our diffusion model approach is also robust against parameter choices as illustrated on the right of Fig. \ref{fig:inferenceNum_compare}. 
The number of diffusion steps $K$ only affects the performance when chosen below a threshold $K<15$, while a fully conditional model with $p=0$ in the training loss \eqref{eq:MPDC_loss} merely causes a small negative effect on the performance. Therefore, this evaluation demonstrates the potential of our diffusion model approach to reliably enable probabilistically, near-globally optimal MPC for control problems beyond the complexity considered in this paper.

\begin{figure}
    \centering
    \def\file{figure/scalable_compare_cost.txt}
    \begin{tikzpicture}
        \begin{semilogxaxis}[ylabel={\sffamily norm. cost},
            xmin=0.0, ymin = 0, xmax = 1000, ymax =1, 
            height =2.9 cm, width=\textwidth/2.6,
            legend pos = north east, legend columns=1, legend entries = {\sffamily our method, \sffamily global optim.},
            name=plot1, ylabel near ticks,
            ylabel shift = -0.15cm, xlabel shift = -0.15cm,
            xticklabels = {},
            yticklabel shift = -0.1cm,
         ]

            \addplot[blue!90!black, mark=*, very thick] table [x = GuessNum, y = OurMethod_Cost]{\file};
            \addplot[orange, mark=*, very thick] table [x = GuessNum, y = MPCMG_Cost]{\file};
            
        \end{semilogxaxis}
        \def\file{figure/scalable_compare_time.txt}
        \begin{loglogaxis}[
            ylabel={\sffamily comp. time [s]~~~~},
            xlabel={\sffamily \# control samples $M$},
            xmin=0.0, ymin = 0, xmax = 1000, ymax =350, 
            height =2.9cm, width=\textwidth/2.6,
            legend pos = north west, legend columns=1, legend entries = {\sffamily our method, \sffamily global optim.},
            name=plot2, ylabel near ticks,
            at=(plot1.east), anchor=east, yshift=-1.6cm,
            ylabel shift = -0.2cm, xlabel shift = -0.15cm,
            yticklabel shift = -0.1cm,
         ]
            \addplot[blue!90!black, mark=*, very thick] table [x = GuessNum, y = OurMethod_Time]{\file};
            \addplot[orange, mark=*, very thick] table [x = GuessNum, y = MPCMG_Time]{\file};
            
        \end{loglogaxis}

        \def\file{figure/horizon_cost.txt}
        \begin{semilogxaxis}[
            xmin=0.0, ymin = 0, xmax = 128, ymax =1, 
            height =2.9cm, width=\textwidth/2.6,
            legend pos = north east, legend columns=1, legend entries = {\sffamily our method, \sffamily global optim.},
            name=plot3, ylabel near ticks,
            at=(plot1.east), anchor=east, xshift=4.88cm,
            ylabel shift = -0.15cm, xlabel shift = -0.15cm,
            yticklabel shift = -0.1cm,
            xticklabels = {},
         ]
            \addplot[blue!90!black, mark=*, very thick] table [x = Horizon, y = OurMethod_medianCost]{\file};
            \addplot[orange, mark=*, very thick] table [x = Horizon, y = MPCMG_medianCost]{\file};
            
            \end{semilogxaxis}

            \def\file{figure/horizon_time.txt}
            \begin{loglogaxis}[
            xlabel={\sffamily \# pred. horizon $H$},
            xmin=0.0, ymin = 0, xmax = 128, ymax =50, 
            height =2.9cm, width=\textwidth/2.6,
            legend pos = north west, legend columns=1, legend entries = {\sffamily our method, \sffamily global optim.},
            name=plot4, ylabel near ticks,
            at=(plot3.east), anchor=east, yshift=-1.6cm,
            ylabel shift = -0.2cm, xlabel shift = -0.15cm,
            yticklabel shift = -0.1cm,
         ]
            \addplot[blue!90!black, mark=*, very thick] table [x = Horizon, y = OurMethod_meanTime]{\file};
            \addplot[orange, mark=*, very thick] table [x = Horizon, y = MPCMG_meanTime]{\file};
            
            \end{loglogaxis}
            \def\file{figure/step_cost.txt}
            \begin{axis}[
            xmin=3, ymin = 0, xmax = 25, ymax =1, 
            height =2.9cm, width=\textwidth/2.6,
            legend pos = north east, legend columns=1, legend entries = {\sffamily our method, \sffamily fully cond.},
            name=plot5, ylabel near ticks,
            at=(plot3.east), anchor=east, xshift=4.88cm,
            ylabel shift = -0.15cm, xlabel shift = -0.15cm,
            xticklabels = {},
            yticklabel shift = -0.1cm,
         ]
            \addplot[blue!90!black, mark=*, very thick] table [x = Step, y = OurMethod]{\file};
            \addplot[orange, mark=*, very thick] table [x = Step, y = FullyCondDDPM]{\file};
            
            \end{axis}

            \def\file{figure/step_time.txt}
            \begin{axis}[
            xlabel={\sffamily \# diff. steps $K$},
            xmin=3, ymin = 0, xmax = 25, ymax = 0.75, 
            height =2.9cm, width=\textwidth/2.6,
            legend pos = north west, legend columns=1, legend entries = {\sffamily our method, \sffamily fully cond.},
            name=plot6, ylabel near ticks,
            at=(plot5.east), anchor=east, yshift=-1.6cm,
            ylabel shift = -0.2cm, xlabel shift = -0.15cm,
            yticklabel shift = -0.1cm,
         ]
            \addplot[blue!90!black, mark=*, very thick] table [x = Step, y = OurMethod]{\file};
            \addplot[orange, mark=*, very thick] table [x = Step, y = FullyCondDDPM]{\file};
            
            \end{axis}
        \end{tikzpicture}
    \vspace{-1.1cm}
\caption{Performance of diffusion model-based and global optimization-based NMPC depending on the number of sampled control sequences $M$ (left), horizon length $H$ (middle), and diffusion steps $K$ (right).
}
\label{fig:inferenceNum_compare}
\end{figure}
 
\vspace{-0.3cm}
\section{Conclusion} \label{conclusion} 
\vspace{-0.15cm}

In this paper, we propose a diffusion model approach to approximate globally optimal nonlinear model predictive control. After training a diffusion model with locally optimal control sequences offline, our approach employs the diffusion model for efficient sampling-based optimization in the online phase. The effectiveness of our diffusion model approach toward near-globally optimal nonlinear model predictive control is demonstrated on the cartpole system. Future work will focus on deriving formal theoretical optimality guarantees for our proposed approach. Moreover, we will investigate the applicability of diffusion models for model predictive control of complex systems, such that we can evaluate it in real-world experiments.

\acks{This work was supported by the European Union’s Horizon Europe
innovation action programme under grant agreement No. 101093822,
“SeaClear2.0”, the DAAD programme Konrad Zuse Schools of Excellence in Artificial Intelligence}, Consolidator Grant ”Safe
data-driven control for human-centric systems” (CO-MAN) of
the European Research Council (ERC) and by a part of NCCR Automation, a National Centre of Competence (or Excellence) in Research, funded by the Swiss National Science Foundation (grant number $\text{51NF40 225155}$)

\bibliography{reference/NumericalSolver, reference/MPC, reference/diffusion, reference/data_generation_(SOP), reference/approxMPC}

\if\isarxiv1
\newpage
\appendix
\section{Data Generation Setting}
There are three dynamic systems a) cart-pole, b) pendubot, and c) double cart-pole in our evaluation. The details of these systems are shown in the \cref{tab:system_detail}. To achieve the swing-up control task for each system, we define quadratic cost functions $\sum_{i=0}^{H-1}\bm{x}_{nl,i}^\top \bm{Q}\bm{x}_{nl,i} + u_i^\top Ru_i$ and terminal costs $\bm{x}_{nl,H}^\top \bm{P}\bm{x}_{nl,H}$ for the nonlinear state $\bm{x}_{nl}$ transformed from original state. The details of the NMPC setting and the parameters for \cref{alg:data_generation} are illustrated in \cref{tab:mpc_detail}.
\begin{table}[]
    \scriptsize
    \renewcommand{\arraystretch}{1.2}
    \centering
    \begin{tabular}{c c c c c c}
        \hline
        & Cart-pole & Pendubot & Double cart-pole \\
        \hline        
        System State
        & $\bm{x} = [x, \dot{x}, \theta, \dot{\theta}]^{\top} \in \mathbb{R}^4$ & $\bm{x} = [\theta_1, \theta_2, \dot{\theta}_1, \dot{\theta}_2]^{\top} \in \mathbb{R}^4$ & $\bm{x} = [x, \dot{x}, \theta_1, \dot{\theta}_1, \theta_2,  \dot{\theta}_2]^{\top} \in \mathbb{R}^6$ \\
        System Input
        & $u \in \mathbb{R}$ & $u \in \mathbb{R}$ & $u \in \mathbb{R}$ \\
        \hline
    \end{tabular}
    \vspace{-3pt}
    \caption{Evaluated systems.}
    \label{tab:system_detail}
\end{table}

\begin{table}[]
    \scriptsize
    \renewcommand{\arraystretch}{1.2}
    \centering
    \begin{tabular}{c c c c c c}
        \hline
        & Cart-pole & Pendubot & Double cart-pole \\
        \hline        
        $\bm{Q}$
        & $diag([0.01,0.01,1000,0.01])$ & $diag([100,100,1,1])$ & $diag([1,1,1000,1,1000,1])$ \\
        $R$
        & 0.001 & 1 & 0.001 \\
        $\bm{P}$
        & $diag([0.01,0.1,1000,0.1])$ & $diag([1000,1000,10,10])$ & $diag([1,1,100,1,100,1])$ \\
        $H$
        & 64 & 256 & 128 \\
        $\bm{x}_{nl}$
        & $[x, \dot{x}, -(\theta-\pi)^2/\pi, \dot{\theta}]^{\top}$ & $[1+\cos(\theta_1), 1-\cos(\theta_2), \dot{\theta}_1, \dot{\theta}_2]^{\top}$ & $[x, \dot{x}, \sin(\theta_1/2), \dot{\theta}_1, \sin(\theta_2/2),  \dot{\theta}_2]^{\top}$ \\
        $\chi$
        & $\{[-3,3],0,[1.8,4.4],0\}$ & $\{0,[-\pi/4,\pi/4],0,0\}$ & $\{[-3,3],0,\pi,0,3.3,0\}$ \\  
        $N_s$
        & 150 & 50 & 50 \\
        $N_T$
        & 50 & 400 & 5 \\
        $N_p$
        & $16$ & $16$ & $6$ \\
        $\bm{\sigma}_d$
        & $0.15$ & $0.05$ & $0.02$ \\
        \hline
    \end{tabular}
    \vspace{-3pt}
    \caption{Detail of data generating.}
    \label{tab:mpc_detail}
\end{table}

\section{Training of Diffusion Models}
Our implementation of the diffusion model is based on the code provided along with the work of  \cite{carvalho2023motion}. We split the dataset into 95\% for training and 5\% for validation, and select the model that achieves the best performance on the validation set for testing. The training details for each system are provided in \cref{tab:train_detail}. Each of the trainings takes around 1 hour under the machine with an AMD EPYC 7542 32 cores processor and NVIDIA Tesla P100.
\begin{table}[]
    \scriptsize
    \renewcommand{\arraystretch}{1.2}
    \centering
    \begin{tabular}{c c c c c c}
        \hline
        & Cart-pole & Pendubot & Double cart-pole \\
        \hline        
        Batch size
        & 4096 & 1024 & 1024 \\
        Epochs
        & 300 & 100 & 1000 \\
        Learning rate
        &$3 \times 10^{-3}$ &$3 \times 10^{-3}$ &$3 \times 10^{-3}$ \\
        Optimizer
        & Adam (\cite{Adam}) & Adam & Adam \\
        Diffusion steps $K$
        & 25 & 25 & 25 \\
        
        \hline
    \end{tabular}
    \vspace{-3pt}
    \caption{Detail of training.}
    \label{tab:train_detail}
\end{table}

\section{Inference Setting}
To validate the performance of our method for each system, we randomly generate $N$ initial states from $\chi$ as shown in \cref{tab:mpc_detail} and test our method in closed-loop control roll-outs over $S$ time steps. $M$ candidates are generated by \cref{alg:mpc_diff} and the first element of the best control sequence is applied to the system. The specific parameters for each dynamical system leading to \cref{fig:performance} are provided in \cref{tab:inference_detail}. 
The ablation study illustrated in \cref{fig:inferenceNum_compare} is based on the same parameters as used for the cart-pole system in other simulations.

\begin{table}[]
    \scriptsize
    \renewcommand{\arraystretch}{1.2}
    \centering
    \begin{tabular}{c c c c c c}
        \hline
        & Cart-pole & Pendubot & Double cart-pole \\
        \hline        
        Samples for the initial state $N$ 
        & 100 & 30 & 20 \\
        Candidates $M$
        & 5 & 30 & 30 \\
        Time step $S$
        &50 &80 &5 \\
        Diffusion steps $K$
        & 25 & 20 & 25 \\
        
        \hline
    \end{tabular}
    \vspace{-3pt}
    \caption{Inference setting for \cref{fig:performance}.}
    \label{tab:inference_detail}
\end{table}

\fi

\end{document}